\documentclass[aps,prl,reprint,noeprint]{revtex4-2}

\usepackage{epsfig}
\usepackage{graphicx}
\usepackage[figuresright]{rotating}

\usepackage{amsfonts}
\usepackage{amsmath}
\usepackage{amssymb}
\usepackage{wasysym}
\PassOptionsToPackage{normalem}{ulem}
\usepackage{ulem}
\usepackage[unicode=true,pdfusetitle,
 bookmarks=true,bookmarksnumbered=false,bookmarksopen=false,
 breaklinks=false,pdfborder={0 0 1},backref=false,colorlinks=false]
 {hyperref}
\hypersetup{colorlinks,linkcolor=blue,citecolor=blue,urlcolor=blue}

\makeatletter

\usepackage{xcolor}
\usepackage{pdfcolmk}
\usepackage{wasysym}

\usepackage{txfonts}
 
\makeatother

\hypersetup{
	pdfborder={0 0 0},
	pdfnewwindow=true,      
	colorlinks=true,       
	linkcolor=blue,          
	citecolor=blue,        
	filecolor=blue,      
	urlcolor=blue,          
}
\usepackage{graphicx} 
\usepackage{amsmath} 

\renewcommand{\vec}[1]{\boldsymbol{#1}}
\renewcommand{\tensor}[1]{\stackrel{\leftrightarrow}{\vec{#1}}}
\newcommand{\me}{\mathrm{e}}
\newcommand{\mi}{\mathrm{i}}

\begin{document}
	
\title{Writing and deleting skyrmions with electric fields in a multiferroic heterostructure}

\author{Chao-Kai Li}
\affiliation{Department of Physics and HKU-UCAS Joint Institute 
for Theoretical and Computational Physics at Hong Kong, 
The University of Hong Kong, Hong Kong, China}
\author{Xu-Ping Yao}
\affiliation{Department of Physics and HKU-UCAS Joint Institute 
for Theoretical and Computational Physics at Hong Kong, 
The University of Hong Kong, Hong Kong, China}
\author{Gang Chen}
\email{gangchen@hku.hk}
\affiliation{Department of Physics and HKU-UCAS Joint Institute 
for Theoretical and Computational Physics at Hong Kong, 
The University of Hong Kong, Hong Kong, China}  

\date{\today}

\begin{abstract}
Magnetic skyrmions are topological spin textures that can be used as information 
carriers for the next-generation information storage and processing. 
The electric-field controlling of skyrmions in such devices is essential 
but remains technologically challenging. Here, using the first-principle 
calculation and the Ginzburg-Landau theory, we propose a reliable 
process for writing and deleting skyrmions by electric fields, 
on the platform of a multiferroic heterostructure, 
particularly the $\text{Cr}_{2}\text{Ge}_{2}\text{Te}_{6} $/$ \text{In}_{2}\text{Se}_{3} $
heterostructure. We show that the electric field controls the electric polarization  
and indirectly influences the antisymmetric Dzyaloshinskii-Moriya interaction (DMI) 
between the magnetic moments. The latter is responsible for the generation  
and removal of the skyrmion spin textures, and we study this mechanism 
by the Ginzburg-Landau analysis. We discuss the real-space Berry curvature, 
topological Hall effects, possible quantum anomalous Hall effect, and other competing magnetic structures. 
These results represent examples of quantum technology and may 
have potential applications in future skyrmionics and the device fabrication. 
\end{abstract}

\maketitle

\emph{Introduction.}---In recent years, magnetic skyrmions have attracted much attention due to the  
spin textures and potential applications in the next-generation information storage 
and processing devices. Magnetic skyrmions are vortexlike topological objects
 in magnetic systems~\cite{Skyrme1962,Bogdanov1989,Bogdanov1994,Bogdanov2001,Binz2006a,Binz2006b,Roessler2006}. 
There has been a significant experimental evidence for their existence 
in condensed matter systems~\cite{Muhlbauer2009,Yu2010,Munzer2010,Heinze2011,Yu2011}. 
Each skyrmion is characterized by a topological invariant 
called the skyrmion number, which is an integer related
to the homotopy mapping from the vector spin space to the 
real space and quantifies the winding of its spin configuration. 
Because of its topological nature, a continuous deformation of 
the spin configuration cannot vary the skyrmion number. 
Thus, magnetic skyrmions are rather robust objects that are topologically protected 
against environmental disturbance, and can be utilized as information 
carriers in future devices like the skyrmion-based racetrack memory~\cite{Fert2013}.

In the potential devices, it is crucial to be able to write and delete the skyrmions. 
It was experimentally demonstrated that skyrmions can be created and destroyed 
by the tunneling current~\cite{Romming2013} or the electric field~\cite{Hsu2017} of 
a scanning tunneling 
microscope (STM) tip, with the latter still involving a tunneling current. Another promising 
way is to control the skyrmions in the multiferroic insulators by 
 electric fields via the electromagnetic coupling~\cite{Seki2012,Mochizuki2015,Wang2020a,Wang2020b}.
This has the advantage of avoiding the inevitable energy dissipation in the current injection
methods, and the spatial position of the skyrmion is less perturbed, which is better 
for a write unit~\cite{Hsu2017}. However, the single-phase multiferroic materials 
are rare because ferromagnetism (ferroelectricity) arises from partially filled (empty) 
$d$ shells of transition metal ions~\cite{Hill2000}.

Since the discovery of graphene~\cite{Novoselov2004}, 
technical advances have made feasible the fabrication of heterostructures 
from different van der Waals (vdW) materials~\cite{Geim2013}
and provide rich degrees of freedom  
to form multifunctional materials.
One crucial ingredient for skyrmions is the antisymmetric exchange interaction, 
known as Dzyaloshinskii-Moriya interaction (DMI)~\cite{Dzyaloshinsky1958,Moriya1960} that
originates from the spin-orbit coupling (SOC). DMI can only exist in systems without
the inversion symmetry. Inversion symmetry breaking is an obvious property of a 
heterostructure; hence, there is usually a finite DMI for the magnetic interaction. 
Moreover, in 3D systems, the skyrmion lattice phase is usually restricted to a narrow region of temperature and external magnetic field. Only with the help of thermal fluctuations can the skyrmion lattice phase be stabilized against the competing conical phase~\cite{Muhlbauer2009}. In contrast, in 2D systems, the competing conical phase is absent for a perpendicular applied magnetic field, bringing about a much more robust skyrmion lattice phase that survives over a wide range of the phase diagram~\cite{Yu2010,Yu2011,Huang2012}. 
For these reasons, vdW heterostructures
 provide a versatile and natural platform for the exploration 
and application of magnetic skyrmions.

\begin{figure}[b]
	\includegraphics[width=8.6cm]{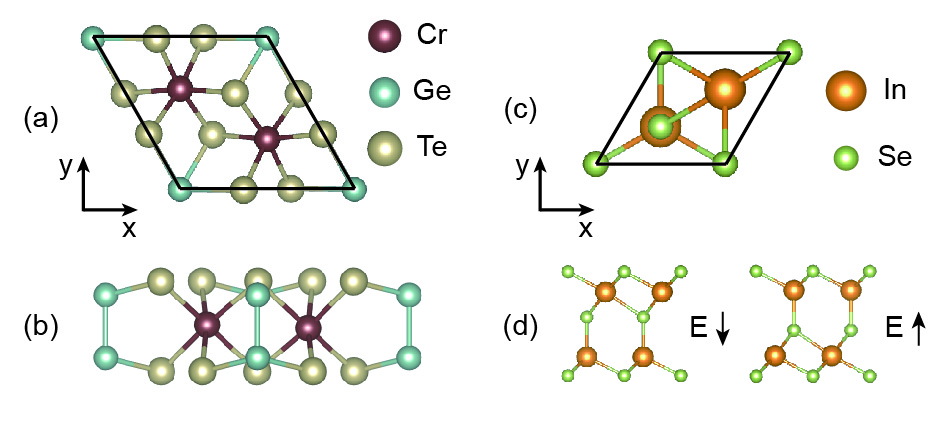}
	\caption{(Color online.) (a) Top view of monolayer $ \text{Cr}_{2}\text{Ge}_{2}\text{Te}_{6} $. 
	(b) Side view of monolayer $ \text{Cr}_{2}\text{Ge}_{2}\text{Te}_{6} $. 
	(c) Top view of monolayer $ \text{In}_{2}\text{Se}_{3} $. 
	(d) Side view of monolayer $ \text{In}_{2}\text{Se}_{3} $.
	The left and right panels show the structures with down and up electric polarization, respectively.
	Crystal structures are drawn by VESTA~\cite{Momma2011}.}
	\label{fig1}
\end{figure}

In this work, we study an electric controllable skyrmion lattice phase in the multiferroic vdW 
heterostructure made of ferromagnetic $\text{Cr}_{2}\text{Ge}_{2}\text{Te}_{6}$ and 
ferroelectric $\text{In}_{2}\text{Se}_{3}$. This is particularly inspired by an interesting recent 
work~\cite{Gong2019} by Gong \textit{et al.}, which proposed that this heterostructure can realize a 
switchable ferromagnet through the control of the electric polarization of $\text{In}_{2}\text{Se}_{3}$. 
By means of first-principles calculations, we find the existence of nonvanishing DMIs between 
the nearest-neighbor $\text{Cr}^{3+} $ spins. The theoretical analysis 
based on the Ginzburg-Landau theory further predicts that in an appropriate external 
magnetic field, the skyrmion lattice phase is more stable than the ferromagnetic phase. 
More substantially, with the switch of the direction of electric polarization of 
$\text{In}_{2}\text{Se}_{3} $, the strength of the DMI undergoes a change so significant 
that the magnetic structure of $ \text{Cr}_{2}\text{Ge}_{2}\text{Te}_{6} $ will switch 
between the topologically distinct skyrmion lattice and the ferromagnetic phase. 
This electric-field-controlled writing and deleting process of skyrmions 
should have potential applications in future skyrmionics devices.

\begin{figure}
	\includegraphics[width=8.6cm]{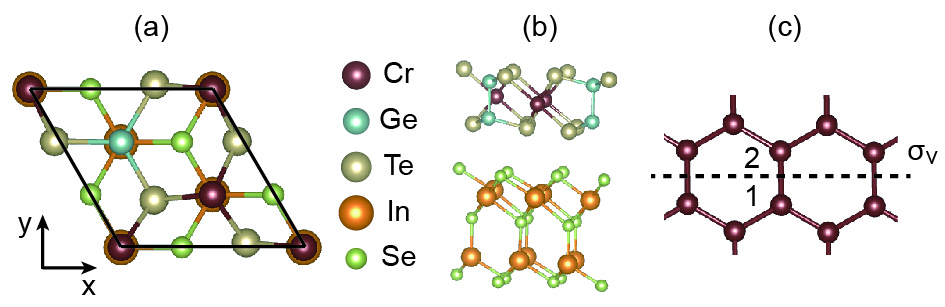}
	\caption{(Color online.) (a) and (b) Top view and side view of the heterostructure 
	$\text{Cr}_{2}\text{Ge}_{2}\text{Te}_{6} $/$ \text{In}_{2}\text{Se}_{3} $, respectively. 
	In (b), the electric polarization of $ \text{In}_{2}\text{Se}_{3} $ is pointing down, as an example. 
	(c) The Cr honeycomb lattice. 
	The interaction between spin 1 and spin 2 is presented in the main text.
	 Crystal structures are drawn by VESTA~\cite{Momma2011}.}
	\label{fig2}
\end{figure}

Bulk $ \text{Cr}_{2}\text{Ge}_{2}\text{Te}_{6}$ is a layered material with the space group 
$ R\bar{3} $~\cite{Carteaux1995}. Below 61~K, it develops a ferromagnetic order with $ \text{Cr}^{3+}$  moments aligned in the $c$ axis. 
The long-range magnetic order has proved to survive in the 2D limit
with an easy-axis anisotropy to counteract the thermal fluctuations~\cite{Gong2017}. 
The structure of monolayer $ \text{Cr}_{2}\text{Ge}_{2}\text{Te}_{6} $ is shown in Figs.~\ref{fig1}(a) and \ref{fig1}(b). The magnetic $ \text{Cr}^{3+} $ ions arrange 
in a honeycomb lattice, as shown in Fig.~\ref{fig2}(c), with each lattice site 
as a center of $ C_3 $ rotational symmetry. Every nearest-neighbor $ \text{Cr}^{3+}$
 ions are related by the inversion symmetry. Thus, the nearest-neighbor DMI is forbidden. 
To introduce nonvanishing DMIs, the inversion symmetry has to be broken. 
This is naturally achieved by manufacturing a heterostructure.

Bulk $ \text{In}_{2}\text{Se}_{3} $ is composed of Se-In-Se-In-Se quintuple layers stacking in the 
$ c $ direction~\cite{Osamura1966}. $ \text{In}_{2}\text{Se}_{3} $ in few-layer forms have recently 
been obtained by mechanical exfoliation and chemical vapor deposition~\cite{Tao2013,Jacobs-Gedrim2014,Lin2013}. It was predicted that the ground state of the monolayer $ \text{In}_{2}\text{Se}_{3} $ 
does not have the middle Se layer equidistant from the two neighboring In layers~\cite{Ding2017}. 
Instead, it is nearer to either the upper or the lower In layer, leading to a spontaneous out-of-plane 
electric polarization whose direction depends on the middle Se layer position. Figures~\ref{fig1}(c) and \ref{fig1}(d) 
depict the monolayer $ \text{In}_{2}\text{Se}_{3} $
 with different polarization directions. The in-plane inversion symmetry is also broken, 
 resulting in an additional in-plane electric polarization. These make $ \text{In}_{2}\text{Se}_{3}$ a 
 2D ferroelectric material, confirmed by later experiments~\cite{Zhou2017,Cui2018}.

\begin{table*}
\begin{ruledtabular}
\begin{tabular}{lccccccccc}
\quad\, Layouts
& $ J_{xx} $ & $ J_{yy} $ & $ J_{zz} $ & $ \Gamma_{xy} $ & $ \Gamma_{xz} $ & $ \Gamma_{yz} $ & $ D_{x} $ & $ D_{y} $ & $ D_{z} $ \\
(case 1) freestanding & -7.99 & -9.13 & -8.85 & 0.00 & 0.09 & 0.00 & 0.00 & 0.00 & 0.00 \\
(case 2) heterostructure ($ \vec{E} \uparrow$) & -11.41 & -12.55 & -12.31 & 0.00 & 0.13 & 0.00 & -0.05 & 0.00 & -0.36 \\
(case 3) heterostructure ($ \vec{E} \downarrow$) & -11.76 & -12.90 & -12.68 & 0.00 & 0.14 & 0.01 & -0.19 & 0.00 & -0.46
\end{tabular}
\end{ruledtabular}
\caption{
The calculated exchange couplings in Eq.~\eqref{eq: tensor J}, in units of meV. 
The spins are normalized with ${| \vec{S} | = 1} $. 
The data for several different structures are listed: 
(1) freestanding monolayer $ \text{Cr}_{2}\text{Ge}_{2}\text{Te}_{6} $; 
(2) $ \text{Cr}_{2}\text{Ge}_{2}\text{Te}_{6} $/$ \text{In}_{2}\text{Se}_{3} $ heterostructure with the upward electric polarization; 
and (3) $ \text{Cr}_{2}\text{Ge}_{2}\text{Te}_{6} $/$ \text{In}_{2}\text{Se}_{3} $ heterostructure with the downward electric polarization.}
\label{tab1}
\end{table*}

The heterostructure that we investigate is comprised of a monolayer 
$\text{Cr}_{2}\text{Ge}_{2}\text{Te}_{6}$ and a monolayer $\text{In}_{2}\text{Se}_{3} $, 
the same as Ref.~[\onlinecite{Gong2019}]. 
The lattice constant of $ \text{Cr}_{2}\text{Ge}_{2}\text{Te}_{6} $ is fixed to the experimental value 
$6.83~\mathring{\text{A}} $. The ${\sqrt{3}\times\sqrt{3}}$ $ \text{In}_{2}\text{Se}_{3} $ 
supercell is strained by $ -4.0\%$ to fit with the $\text{Cr}_{2}\text{Ge}_{2}\text{Te}_{6}$ primitive cell. 
Specifically, the interfacial Te atoms lie above the hollow sites of the interfacial 
$\text{In}_{2}\text{Se}_{3}$ hexagon, as shown in Figs.~\ref{fig2}(a) and \ref{fig2}(b). 
This is the configuration with the lowest energy~\cite{Gong2019}. 
A vacuum layer of $20~\mathring{\text{A}}$ is introduced in the slab model to 
avoid the artificial interlayer interactions between the periodic images. 
The coordinates of the atoms inside the unit cell are relaxed.

The simulations are done within the framework of density functional theory~\cite{Hohenberg1964,Kohn1965}, 
using the plane wave basis sets and pseudopotential method as implemented in the {\sc Quantum ESPRESSO} 
package~\cite{QE-2009,QE-2017}. The energy cutoff for the plane wave basis is 72~Ry. 
The projector augmented-wave~\cite{Blochl1994} pseudopotentials 
in the 
pslibrary~\cite{DalCorso2014,pslibrary} are used. 
The Brillouin zone is sampled by a ${6 \times 6 \times 1 }$ grid in structural optimizations. 
The convergence threshold for force is $10^{-3} $~Ry/bohr. The Perdew-Burke-Ernzerhof 
exchange-correlation functional~\cite{Perdew1996} is used together with Grimme's D2
 parameterization~\cite{Grimme2006} of vdW correction to get a reliable interlayer distance. 
 In the energy calculations, the local spin density approximation (LSDA)~\cite{Perdew1981} is adopted. 
 Fully relativistic pseudopotentials are employed to account for the SOC. On-site Hubbard $U$ 
 value is set to $0.5$~eV for Cr $d$ orbitals~\cite{Gong2017}, which are orthogonalized using 
 L\"owdin’s method~\cite{Mayer2002}. Four-state energy-mapping method~\cite{Xiang2011,Xiang2013} 
 is used to obtain the magnetic couplings. A ${1 \times \sqrt{3}}$ supercell is adopted 
 in the energy calculations to reduce the spin interaction with their periodic images, 
 and the corresponding Brillouin zone is sampled by a ${8 \times 5 \times 1}$ grid. 
 The convergence threshold of total energy for self-consistency is $10^{-7} $~Ry.

The optimized interlayer distances are $3.15~\mathring{\text{A}} $ and $3.10~\mathring{\text{A}} $ 
for the electric polarization of $\text{In}_{2}\text{Se}_{3}$ pointing up and down, respectively. 
The total energy per unit cell with down polarization is lower than that with up polarization by 0.1~eV 
with SOC included and 0.06~eV without SOC. The magnetic moment of each $ \text{Cr}^{3+} $ 
ion is about $ 3\mu_B $ for spin 3/2. These results are similar with the earlier report~\cite{Gong2019}.
We adopt the generic Hamiltonian
\begin{equation}
H_{12}=\vec S_{1}\cdot\tensor J\cdot\vec S_{2}
\end{equation}
to describe the spin interaction between the Cr spins at site 1 and site 2 in Fig.~\ref{fig2}(c). 
The exchange matrix has the components
\begin{equation}
\label{eq: tensor J}
\tensor J=\left[\begin{array}{ccc}
J_{xx} & {\Gamma_{xy}+D_{z} }& {\Gamma_{xz}-D_{y}}\\
{\Gamma_{xy}-D_{z}} & J_{yy} & { \Gamma_{yz}+D_{x}}\\
{\Gamma_{xz}+D_{y}} &{ \Gamma_{yz}-D_{x}} & J_{zz}
\end{array}\right]
\end{equation}
in Cartesian coordinates. Here, the $ J $'s are the Heisenberg interactions, 
the $\Gamma $'s are the off-diagonal pseudodipolar interactions, and the 
$D$'s are the DMIs. The Hamiltonian of other nearest-neighbor Cr sites can 
be deduced from $H_{12} $. Our results of the parameters in Eq.~(\ref{eq: tensor J}) 
for normalized spin vector, ${| \vec{S} | = 1}$, are listed in Table~\ref{tab1}.

First, we analyze the case for a freestanding $\text{Cr}_{2}\text{Ge}_{2}\text{Te}_{6} $. 
As mentioned earlier, the nearest-neighbor Cr sites are related by inversion symmetry, 
leading to ${{\boldsymbol D} = {\boldsymbol 0}}$. Moreover, there is an approximate 
symmetry of the mirror plane $\sigma_{v}$ perpendicular to the line joining the 
nearest-neighbor Cr sites, as Fig.~\ref{fig1}(c) shows, 
although it is slightly broken~\cite{Carteaux1995}. In the following, 
we take this symmetry into consideration, and the point group becomes $C_{3v}$. 
Then, we have the symmetry restriction ${\Gamma_{xy}, \Gamma_{yz} \approx 0}$. 
Indeed, we find a vanishingly small $ \Gamma_{xy} $ and $ \Gamma_{yz} $, 
as can be seen from case 1 in Table~\ref{tab1}, validating our assumption.

When the monolayer $\text{Cr}_{2}\text{Ge}_{2}\text{Te}_{6}$ is placed on top of the ferroelectric 
monolayer $\text{In}_{2}\text{Se}_{3}$ to form a multiferroic heterostructure~\cite{Gong2019}, 
the naturally broken inversion symmetry generates finite DMIs. For the configuration shown in 
Fig.~\ref{fig2}(a), the aforementioned approximate mirror plane symmetry $\sigma_{v}$ still holds
and imposes the restriction ${ \Gamma_{xy}, \Gamma_{yz}, D_{y} \approx 0 }$. 
Hence, the introduction of $\text{In}_{2}\text{Se}_{3} $ leads to nonzero $D_{x}$ and $D_{z}$, 
and this is supported by the numerical results in case 2 and case 3 of Table~\ref{tab1}. 
It is quite remarkable that the direction of the electric polarization of the substrate 
$ \text{In}_{2}\text{Se}_{3} $ has a significant influence on the magnitude of $D_{x}$. We
find $D_{x}$ is about four times larger with a downward electric polarization.

In contrast to the Heisenberg interactions, the DMIs necessarily favor noncollinear 
spin alignments. This is an important ingredient for the skyrmion formation. 
To discuss the possible skyrmion lattice phase, we employ the Ginzburg-Landau theory. 
For the $C_{3v}$ symmetry, the general free-energy functional of the spin distribution 
$\vec{S}(\vec{r})$ under a magnetic field $B$ along $z$ direction is
\begin{eqnarray}
\label{eq: continuum model}
 F&=&\int d^{2}\vec r\Bigg\{\frac{\tilde{J}_{1}}{2}\left[\left(\partial_{x}S_{x}\right)^{2}
 +\left(\partial_{y}S_{y}\right)^{2}\right]
+\frac{\tilde{J}_{2}}{2}\left[\left(\partial_{x}S_{y}\right)^{2}
+\left(\partial_{y}S_{x}\right)^{2}\right] \nonumber\\
&&+\frac{\tilde{J}_{3}}{2}
\left[\left(\partial_{x}S_{z}\right)^{2}+\left(\partial_{y}S_{z}\right)^{2}\right]
+\left(\tilde{J}_{1}-\tilde{J}_{2}\right)\partial_{x}S_{x}\partial_{y}S_{y}
\nonumber 
\\
 &&
+\tilde{\Gamma}\left(\partial_{x}S_{x}\partial_{x}S_{z}
-2\partial_{x}S_{y}\partial_{y}S_{z}-\partial_{y}S_{x}\partial_{y}S_{z}\right) \nonumber\\
&&+2\tilde{D}\left(S_{x}\partial_{x}S_{z}+S_{y}\partial_{y}S_{z}\right)
+\tilde{A}S_{z}^{2}+\tilde{C}S^{2}-BS_{z} \Bigg\}.
\end{eqnarray}
Only terms linear and quadratic in $\vec{S}(\vec{r}) $ are considered at this stage. 
The parameters can be related to the microscopic ones in Table~\ref{tab1} 
by the Taylor expansion
\begin{equation}
{ S_{\mu}(\vec r+\vec d) } \approx 
S_{\mu}(\vec r)+\vec d\cdot\vec{\nabla}
S_{\mu}(\vec r)+\frac{1}{2}(\vec d\cdot\vec{\nabla})^{2}S_{\mu}(\vec r) ,
\end{equation}
for $ {\mu = x, y, z} $. The results for the coefficients in the free energy are listed in the Appendix. 
By Fourier transformation
${\vec S(\vec r)={\left(2\pi\right)^{-2}} 
\int d^2{\vec q}\, \tilde{\vec S}(\vec q)\me^{\mi\vec q\cdot\vec r} }$, 
the free energy becomes
\begin{equation}
F=\int\frac{d^{2}\vec q}{\left(2\pi\right)^{2}}\sum_{\mu,\nu=1}^{3}\tilde{S}_{\mu}
(\vec q)T_{\mu\nu}(\vec q)\tilde{S}_{\nu}(-\vec q)
-B\tilde{S}_{z}(\vec 0),
\end{equation}
where the expression for $T(\vec q)$ is given in the Appendix.
We have made a division 
${T(\vec q)=T_{0}(\vec q)+T'(\vec q)}$,
$T_{0}(\vec q) $ is obtained from $ T\left(\vec q\right)$ by 
setting ${\tilde{J}_{1}=\tilde{J}_{3}=\tilde{J}}$, ${\tilde{\Gamma}=0}$, 
 and ${\tilde{A}=0}$, and we treat ${T'\left(\vec q\right)}$ perturbatively. 
 This is reasonable for the parameters in Table~\ref{tab1}.
  Letting ${q_{x}=Q\cos\theta}$ and ${q_{y}=Q\sin\theta}$, 
  the lowest eigenvalue of $ T_0(\vec q) $ is minimized 
  to $ -\tilde{D}^{2}/2\tilde{J} $ when the momentum ${Q=-\tilde{D}/\tilde{J}}$.
For our case, ${\tilde{D}<0}$. The corresponding eigenvector is
\begin{equation}
\label{eq: eigenvector}
{\vec{e}(\theta)=\left(  - \frac{\mi \cos\theta,}{\sqrt{2}} - \frac{\mi\sin\theta}{\sqrt{2}},
\frac{1}{\sqrt{2}}\right)^T}.
\end{equation}

\begin{figure}
	\includegraphics[width=8.7cm]{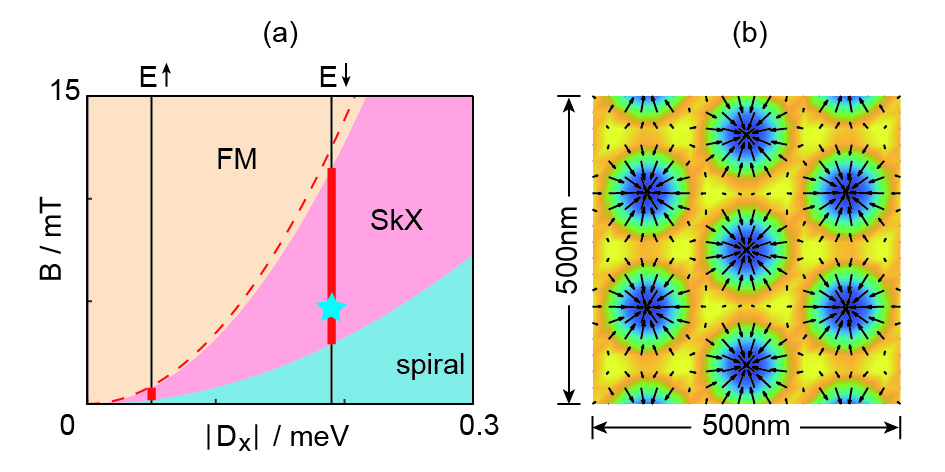}
	\caption{(Color online.) (a) $ B $-$ D_x $ phase diagram. 
	$B$ is the magnetic field and $D_x$ is the strength 
	of the DMI. Ferromagnetic (FM), skyrmion lattice (SkX), 
	and single spiral phases are denoted by yellow, red, and blue regions, respectively. 
	The $D_x$ for up and down electric fields are shown by two black vertical lines, 
	where the segments in the SkX phase are highlighted by red thick lines. 
	The yellow region below the red dashed line has a noncollinear spin texture 
	whose skyrmion number is zero and is smoothly connected to a FM.
(b) A typical skyrmion lattice configuration corresponding to the blue star 
in (a). The in-plane spin components are in the black arrows. 
The out-of-plane spin components are encoded by colors, 
with the blue regions pointing downward and the yellow regions pointing upward.}
\label{fig3}
\end{figure}

The lowest eigenvalue of $T_{0}\left(\vec q\right)$ is independent of $\theta$, 
and thus the minimal free energy is degenerate for arbitrary orientation of the 
wave vector $\vec{q}$ with ${{q} = Q}$. This is a contour degeneracy 
for the free energy very much like the one in the spiral spin liquid~\cite{2020arXiv201103007Y}.
This continuous degeneracy is lifted by the second-order perturbation of 
$ T'(\vec q) $, that produces a $ \theta $-dependent term
${- \frac{1}{16}  \tilde{D}^{2} \tilde{\Gamma}^{2} {\tilde{J}^{-3} (\tilde{J}_{2}-\tilde{J} )
(\tilde{J}_{2}+\tilde{J})^{-1}} \cos 6\theta }$. 
Therefore, the lowest eigenvalue of $T(\vec q)$ 
will get its minimum value at six discrete $ \theta_{i} $, $ i=1,\ldots,6 $.
Having this six-fold degeneracy, we further construct an ansatz of the 
spin configuration from three degenerate spirals and a ferromagnetic $z$ component
\begin{equation}
\label{eq: ansatz}
\vec S(\vec r)=\frac{1}{\sqrt{2}}\sum_{i=1}^{6}
\phi_{i}\me^{\mi\vec q_{i}\cdot\vec r}\vec e_{i}
+\phi_{0}\vec e_{z},
\end{equation}
where the $\vec{e}_{i} $'s are the eigenvectors in Eq.~(\ref{eq: eigenvector}) 
for $\theta_{i} $, and the $\phi_{i}$'s are the corresponding order parameters. 
Moreover, we have ${\theta_{2}=\theta_{1}+2\pi/3}$, ${\theta_{3}=\theta_{1}+4\pi/3}$, 
${\theta_{i+3}=\theta_{i}+\pi}$ (${i=1,2,3}$), and ${\phi_{i+3}^{}=\phi_{i}^{\ast}}$. 
The ansatz is subject to the soft-spin-like constraint that the spatial average of 
the spin vector is normalized~\cite{PhysRevB.83.184406}
\begin{equation}
\label{eq: constraint}
\left\langle \Big|\vec S(\vec r)\Big|^{2}\right\rangle =\sum_{i=1}^{3}\left|\phi_{i}\right|^{2}+\phi_{0}^{2}=1.
\end{equation}
We find that the free-energy density for this spin configuration is
${f=\frac{1}{2} {\tilde{D}^{2}}{\tilde{J}}^{-1}\phi_{0}^{2}-B\phi_{0}^{}
- \frac{1}{2}{\tilde{D}^{2}}{\tilde{J}}^{-1}}$, 
where the wave vector ${q=Q}$ and the constraint in Eq.~(\ref{eq: constraint}) are used. 
The free energy is minimized at ${\phi_{0}={B \tilde{J} }{\tilde{D}^{-2}}}$. 
However, there is still a degeneracy for different values of $\phi_{i} $ ($ {i=1,2,3 }$) 
as long as the constraint Eq.~(\ref{eq: constraint}) is honored. That is, 
the energy of a skyrmion lattice solution for which all the three $\phi_{i} $'s 
are equal in magnitude is degenerate with the single spiral solution for which 
only one of the three $\phi_{i} $'s are nonzero. This degeneracy remains 
after including the $ T'(\vec q) $ correction since it is still quadratic 
in $\vec S(\vec r) $.

The skyrmion lattice is at last stabilized by the quartic terms 
in the free energy~\cite{Muhlbauer2009}. For simplicity, 
we only include the leading isotropic term
${\Delta F=\int d^2{\boldsymbol r} \, | {\boldsymbol S}(\vec r)| ^{4}}$. 
The coefficient, that is positive, is dropped because it does not matter in the following discussion. 
For the spin configuration in Eq.~(\ref{eq: ansatz}), the addition to the free-energy density 
due to the quartic term is
\begin{eqnarray}
\Delta f&=&\phi_{0}^{4}+\left(4\phi_{0}^{2}+\sum_{i=1}^{3}\left|\phi_{i}\right|^{2}\right)\left(\sum_{i=1}^{3}\left|\phi_{i}\right|^{2}\right) 
+ \frac{5}{8}  \sum_{i\neq j} |\phi_i|^2 |\phi_j|^2
\nonumber 
\\ 
&&
+\frac{9}{2}\phi_{0}\left(\phi_{1}\phi_{2}\phi_{3}+\phi_{1}^{*}\phi_{2}^{*}\phi_{3}^{*}\right).
\end{eqnarray}
For a skyrmion lattice solution, we have 
\begin{eqnarray}
\left\{ 
\begin{array}{l}
\left|\phi_{1}\right|=\left|\phi_{2}\right|=\left|\phi_{3}\right|=[{\frac{1}{3}  ( 1-\phi_{0}^{2} )}]^{\frac{1}{2}}, \\
\Delta f_{\text{SkX}}=\phi_{0}^{4}+\left(1+3\phi_{0}^{2}\right)\left(1-\phi_{0}^{2}\right)+\frac{5}{12}\left(1-\phi_{0}^{2}\right)^{2} \\
\quad\quad\quad  \,\,+\sqrt{3}\phi_{0}\left(1-\phi_{0}^{2}\right)^{3/2}\cos\left(\alpha_{1}+\alpha_{2}+\alpha_{3}\right),
\end{array}
\right.
\end{eqnarray}
where $\alpha_{i}$ is the phase of $\phi_{i} $ ($ i=1,2,3 $). 
For a single-$q$ spin spiral, we instead have
\begin{eqnarray}
\left\{
\begin{array}{l}
{\left|\phi_{1}\right|=({1-\phi_{0}^{2}})^{\frac{1}{2}}},\quad\quad \phi_{2}=\phi_{3}=0, \\
\Delta f_{\text{spiral}}=\phi_{0}^{4}+\left(1+3\phi_{0}^{2}\right)\left(1-\phi_{0}^{2}\right).
\end{array}
\right.
\end{eqnarray}
The condition of ${\Delta f_{\text{SkX}}<\Delta f_{\text{spiral}}}$ demands that  
\begin{equation}
\cos\left(\alpha_{1}+\alpha_{2}+\alpha_{3}\right)<-\frac{5}{12} 
\Big[ {\frac{1}{3}\left({\phi_{0}^{-2}}-1\right)} \Big]^{ \frac{1}{2}}.
\end{equation}
The solution of the above inequality exists when $\phi_0$ satisfies 
${\frac{5}{\sqrt{457}}<\phi_{0}<1}$. With ${\phi_0 = B\tilde{J}\tilde{D}^{-2}}$, 
the range of $B$ in favor of the formation of the skyrmion lattice is given by
\begin{equation}
\label{eq: B range}
\frac{5\tilde{D}^{2}  \tilde{J}^{-1} }{\sqrt{457} }  < {B} < {\tilde{D}^{2} \tilde{J}^{-1}} . 
\end{equation}

From the above analysis, it can be seen that when the system is put under an external magnetic field, 
a skyrmion lattice will emerge when the field strength is in the range of Eq.~(\ref{eq: B range}). 
The $ B $-$ D_x $ phase diagram is shown in Fig.~\ref{fig3}, 
in which we have assumed an averaged $ \tilde{J} $ of up and down electric polarization, 
and the magnetic moment of a $ \text{Cr}^{3+} $ ion is set to $ 3\mu_B $. 
As the direction of the electric polarization of $ \text{In}_{2}\text{Se}_{3} $ 
is altered from upward to downward, there is a significant change of the magnitude 
of $ D_{x} $ as large as four times. The $ D_x $ for the two electric polarization 
directions is shown by two vertical lines, and the range of magnetic field in which 
skyrmion lattice can exist is highlighted by red segments. The two red segments 
do not overlap on the $ B $ axis, which means the skyrmion lattice can be created 
and destroyed by two different electric field directions without changing the magnetic field.

\emph{Discussion.}---The size of the skyrmion lattice is determined by the magnitude 
of the wave vector that is proportional to the DMI. In Fig.~\ref{fig3}, we depict a typical 
skyrmion lattice configuration for a downward electric field 
 and ${ B=4 }$~mT. The corresponding phase point is indicated by the blue star 
 in Fig.~\ref{fig3}(a). The skyrmions are of the hedgehog-type. 
 The skyrmion lattice can be detected by various experimental techniques, 
 such as neutron scattering~\cite{Muhlbauer2009,Munzer2010}, 
 Lorentz transmission electron microscopy~\cite{Yu2010,Yu2011,Yu2012}, 
 spin-resolved scanning tunneling microscopy~\cite{Heinze2011}, 
 and topological Hall effect measurements~\cite{Neubauer2009,Lee2009,PhysRevLett.119.176809}.
Topological Hall effect arises from the real-space Berry curvature 
due to the noncollinear spin configuration of the skyrmion lattice that functions as an 
orbital magnetic field or flux for the conduction electrons. It is well-known that, 
the polar heterostructure often confines a conducting 2D electron gas at the interface.
Since a rather large effective magnetic field could be realized by the skyrmion lattice
and this skyrmion lattice can be tuned from our study here, 
it is feasible that, once the conduction electron density is commensurate with the 
magnetic flux of the skyrmion lattice, {\sl quantum anomalous Hall effect} (QAHE)
could emerge. This QAHE would be fundamentally different from the one that was 
realized in the Cr-doped Bi$_2$Se$_3$ thin films~\cite{2013Sci...340..167C}. 
Over there, it is the exchange field
from the Cr ferromagnetic order that renders a mass-gap to the Dirac fermion
on the surface of topological insulator, and the resulting state is 
a Chern insulator with the Chern number ${C=1}$ for the valence band. The proposed 
QAHE here is making use of the effective orbital magnetic field from the noncollinear
spin textures, and is probably closer in spirit to the conventional Landau level
integer quantum Hall effect. It is also reasonable to envision the possibility 
of a fractional QAHE once the electron correlation is included.

Despite the energy difference between the upward and downward electric polarization 
phases, as well as the ferromagnetic (FM) and skyrmion lattice (SkX) phases, there is an 
energy barrier between them. The barrier of the most effective kinetic path between the
two electric polarization directions was 
calculated to be 0.066~eV per unit cell~\cite{Ding2017}. There are 
experiments~\cite{Zhou2017,Cui2018} demonstrating that a bias voltage of several 
volts between the piezoresponse force microscopy tip and the 
$ \text{In}_{2}\text{Se}_{3} $ thin film can switch the electric polarization direction. 
Therefore, the required electric field 
strength is accessible. The energy barrier between the FM and SkX 
phases and the spin dynamics related to the phase transition will be left to 
future works. Even if the barrier were high, we think the phase transition can be 
induced by a suitably applied magnetic field disturbance, which is superposed on the 
perpendicular magnetic field required by the stabilization of SkX phase. For example, 
Ref.~[\onlinecite{Flovik2017}] simulates the 
creation of skyrmions via the application of a tilted magnetic field pulse. The tilted 
magnetic field pulse excites spin waves in the FM phase, making the system energy higher 
than the barrier, so that the system will not get stuck in some metastable state. 
Then, the system can relax to the lower energy SkX phase. The transition 
from SkX to FM phase can also be induced by such disturbance.

\emph{Conclusions.}---In summary, we have proposed an electric-field-controlled writing and 
deleting scheme of the magnetic skyrmions in the multiferroic vdW heterostructure 
$\text{Cr}_{2}\text{Ge}_{2}\text{Te}_{6} $/$ \text{In}_{2}\text{Se}_{3}$ 
through electromagnetic coupling. The inversion symmetry breaking of 
the interface leads to nonvanishing DMIs between neighboring Cr local moments. 
Because of the introduction of the DMIs, a skyrmion lattice emerges in 
$\text{Cr}_{2}\text{Ge}_{2}\text{Te}_{6} $ with an appropriate magnetic field, 
and depends on the strength of the DMIs. 
The strength of the DMI is very sensitive to the direction of the electric polarization 
of the ferroelectric $ \text{In}_{2}\text{Se}_{3} $, providing a natural scheme
for the quantum controlling. Our calculations show that it is 
possible to create and destroy the skyrmion lattice phase by changing 
the direction of the electric polarization. Our findings may have a potential 
application in the future quantum technology 
such as the next-generation information storage and processing devices.

\emph{Note added.}---Upon the acceptance of the current
manuscript, we learned that a recent experimental work on
multiferroic heterostructure consists of Pt/Co/Ta magnetic
multilayer and ferroelectric $ \text{Pb(Mg}_{1/3}\text{Nb}_{2/3}\text{)}_{0.7}\text{Ti}_{0.3}\text{O}_{3} $ was
published in~[\onlinecite{Ba2021}] where the identical phenomena were experimentally realized.

\begin{acknowledgments}
\emph{Acknowledgments.}---We acknowledge Professor Xiang Zhang for communication. 
This work is supported by the Ministry of Science and Technology of China 
with Grant Nos. 2018YFE0103200, 2016YFA0300500, and 2016YFA0301001, by Shanghai Municipal Science and Technology Major Project with Grant No.2019SHZDZX04, and by the Research Grants Council of Hong Kong with General Research Fund Grant No.17303819. The calculation of this work was performed on TianHe-2. Thanks for the support of National Supercomputing Center in Guangzhou (NSCC-GZ).
\end{acknowledgments}

\appendix

\section{The coefficients in the Ginzburg-Landau theory}

The coefficients in the Ginzburg-Landau theory are given as 
\begin{eqnarray}
&&\tilde{J}_{1}=-\frac{J_{xx}+3J_{yy}}{4\sqrt{3}}, \quad \tilde{J}_{2}=-\frac{3J_{xx}+J_{yy}}{4\sqrt{3}},\\
&&\tilde{J}_{3}=-\frac{J_{zz}}{\sqrt{3}}, \quad \tilde{\Gamma}=\frac{\Gamma_{xz}}{2\sqrt{3}}, \quad \tilde{D}=\frac{D_{x}}{a},\\
&&\tilde{A}=\frac{\sqrt{3}}{a^{2}}\left(2J_{zz}-J_{xx}-J_{yy}+\frac{4}{3}A\right),\\
&&\tilde{C}=\frac{\sqrt{3}}{a^{2}}\left(J_{xx}+J_{yy}\right).
\end{eqnarray}
In $ \tilde{A} $, we have included a contribution from single ion anisotropy $ A $.
$T\left(\vec q\right)$ is given as 
\begin{align}
&T\left(\vec q\right)=\tilde{C}I_{3\times3} + \nonumber \\
&\left[\begin{array}{ccc}
\frac{\tilde{J}_{1}}{2}q_{x}^{2}+\frac{\tilde{J}_{2}}{2}q_{y}^{2} & \frac{\tilde{J}_{1}-\tilde{J}_{2}}{2}q_{x}q_{y} & \frac{\tilde{\Gamma}}{2}\left(q_{x}^{2}-q_{y}^{2}\right)-\mi\tilde{D}q_{x}\\
\frac{\tilde{J}_{1}-\tilde{J}_{2}}{2}q_{x}q_{y} & \frac{\tilde{J}_{1}}{2}q_{y}^{2}+\frac{\tilde{J}_{2}}{2}q_{x}^{2} & -\tilde{\Gamma}q_{x}q_{y}-\mi\tilde{D}q_{y}\\
\frac{\tilde{\Gamma}}{2}\left(q_{x}^{2}-q_{y}^{2}\right)+\mi\tilde{D}q_{x} & -\tilde{\Gamma}q_{x}q_{y}+\mi\tilde{D}q_{y} & \frac{\tilde{J}_{3}}{2}\left(q_{x}^{2}+q_{y}^{2}\right)+\tilde{A},
\end{array}\right],
\end{align}

\bibliography{refs}

\end{document}